\documentstyle[aps,twocolumn]{revtex}

\begin{document}
\author{T. P. Cheng$^2$ and Ling-Fong Li$^1$}
\address{$^1$Department of Physics, Carnegie-Mellon University, Pittsburgh, PA
15213\\
$^2$Department of Physics and Astronomy, University of Missouri, St. Louis,
MO 63121}
\title{{\bf Flavor and Spin Contents of the Nucleon} {\bf in the Quark Model
with
Chiral Symmetry}}
\date{CMU-HEP94--30, hep-ph/9410345}
\maketitle

\begin{abstract}
A simple calculation in the framework of the chiral quark theory of Manohar
and Georgi yields results that can account for many of the ''failures'' of
the naive quark model: significant strange quark content in the nucleon as
indicated by the value of $\sigma _{\pi N},$ the $\overline{u}$-$\overline{d}
$ asymmetry in the nucleon as measured by the deviation from Gottfried sum
rule and by the Drell-Yan process, as well as the various quark
contributions to the nucleon spin as measured by the deep inelastic
polarized lepton-nucleon scatterings.
\end{abstract}


One of the outstanding problems in non-perturbative QCD is to understand, at
a more fundamental level, the successes and failures of the non-relativistic
quark model. An important step in that direction was taken years ago by
Manohar and Georgi\cite{mgtheor} when they presented their chiral quark
model with an effective Lagrangian for quarks, gluons and Goldstone bosons
in the region between the chiral symmetry breaking and the confinement
scales. They demonstrated how the successes of the simple quark model could
be understood in this framework, which naturally suggests a chiral symmetry
breaking scale $\Lambda _{\chi SB}$ $\simeq 1\,GeV$, significantly higher
than the QCD confinement scale. And it also allows for the possibility for a
much reduced (effective) strong coupling $\alpha _S$, thus the result of
hadrons as weakly bound states of constituent quarks. A meaningful
calculation of the magnetic moments of the baryons (with great success) can
also be carried out, etc.

In this note we shall show that this framework, with a simple extension, can
also remedy many of the ''experimental failures'' of the simple quark model.
It has been known for a long time that the measured value of the pion
nucleon sigma term $\sigma _{\pi N}\,$ indicates a significant strange quark
presence in the nucleon\cite{tpsigma}. This failure of the simple quark
model is further high-lighted when the deep inelastic polarized muon-proton
scattering measurements made by EMC\cite{emc} indicated a significant
contribution to the proton spin by the strange quarks in the sea\cite{ejsum}%
 Then came along the NMC\cite{nmc} result showing that the Gottfried sum
rule\cite{gsr} is not satisfied experimentally, indicating a nucleon sea
that is quite asymmetric with respect to its $\overline{u}$ and $\overline{d}
$ quark contents. This basic piece of physics is now confirmed by a
dedicated Drell-Yan experiment colliding protons on proton and on neutron
targets\cite{na51}.

An important feature of the chiral quark model is that the internal gluon
effects can be small compared to the internal Goldstone bosons and quarks.
As we shall demonstrate, this picture can account for, in terms of two
parameters, the broad pattern of the observed flavor and spin contents of
the nucleon. In the chiral quark model, the dominant process is the
fluctuation of quark into quark plus a Nambu-Goldstone boson. This basic
interaction causes a modification of the spin content because a quark can
change its helicity by emitting a spin zero meson, Fig. 1(a). It causes a
modification of the flavor content because the Goldstone boson will in turn
fluctuate into a quark-antiquark pair, Fig. 1(b). Instead of using the
parton evolution equation of the chiral quark theory, as has been carried
out by Eichten, Hinchliffe, and Quigg\cite{ehq}, we will illustrate the
basic physical mechanism with a schematic calculation, as was also
considered by these authors.

In the absence of interactions, proton is made up of two $u$ quarks and one $%
d$ quark. We now calculate proton's flavor content after any one of these
quarks emits a quark-antiquark pair via Goldstone bosons, which have
interaction vertices: $\,\,\,{\cal L}_I=g_8\overline{q}\widehat{\phi }q,$%
\[
\,\widehat{\phi }=\sum_{i=1}^8\lambda _i\phi _i=\left(
\begin{array}{ccc}
\frac{\pi ^o}{\sqrt{2}}+\frac \eta {\sqrt{6}} & \pi ^{+} & K^{+} \\
\pi ^{-} & -\frac{\pi ^o}{\sqrt{2}}+\frac \eta {\sqrt{6}} & K^o \\
K^{-} & \overline{K^o} & -\frac{2\eta }{\sqrt{6}}
\end{array}
\right) ,
\]
$q=(u,d,s),$ and the $\lambda ^{\prime }s$ are the Gell-Mann matrices. We
have suppressed all the spacetime structure and have only displayed the
flavor content of the coupling. The probability amplitudes of meson emission
from an $u$ quark to various meson-quark states (as given by the
coefficients in front of the channel names) are
\begin{equation}
\text{\thinspace }\Psi \left( u\right) =g_8\left( \pi ^{+}d+K^{+}s+\frac 1{%
\sqrt{2}}\pi ^ou+\frac 1{\sqrt{6}}\eta u\right) .  \label{psi-u}
\end{equation}
{}From this, we deduce that $1-\frac 83a$ is the probability of no meson
emission with $a$ being the probability of emitting a $\pi ^{+}$ or a $%
K^{+}:\;a\propto \left| g_8\right| ^2.$ By substituting the quark content of
the mesons into the above equation, one obtains the proton's flavor
composition (after one interaction):
\[
\left( 1-\frac 83a\right) \left( 2u+d\right) +2\left| \Psi \left( u\right)
\right| ^2+\left| \Psi \left( d\right) \right| ^2
\]
with
\begin{equation}
\begin{array}{c}
\left| \Psi \left( u\right) \right| ^2=\frac 29\left[ 14u+2\overline{u}%
+5\left( d+\overline{d}+s+\overline{s}\right) \right] a, \\
\left| \Psi \left( d\right) \right| ^2=\frac 29\left[ 14d+2\overline{d}%
+5\left( u+\overline{u}+s+\overline{s}\right) \right] a\,.
\end{array}
\label{psi2}
\end{equation}
In this picture the strange quarks in the sea is brought about by the
fluctuation of the valence $u$ and $d$ quarks into the $s$-quark-containing
mesons: $K$'s and $\eta .$ An asymmetry develops between $\overline{u}$ and $%
\overline{d}$ distributions because there is an initial $u$-$d$ asymmetry in
the valence quarks and the $u\,$quark cannot fluctuate into $\pi ^{-}$
(hence a final state containing $\overline{u}$) while the $d$ quark cannot
fluctuate into $\pi ^{+},$ etc.

Since the $1/N_c$ expansion, $N_c$ being the number of colors, is thought to
be a useful approximation scheme for QCD, it will be worthwhile to see what
will it teach us here. The leading contribution comes from the planar
diagrams. At this order, there are {\it nine} Goldstone bosons, including
the unmixed diagonal channels: $\overline{u}u,\,\,\overline{d}d$ and $%
\overline{s}s,$ all with the same couplings. If we express this in the
language of SU(3), besides the octet there is also the singlet, with the
singlet Yukawa coupling being equal to the octet coupling: $g_0=g_8.$ When
this ninth singlet $\eta ^{\prime }$ meson (here it is not the physical $%
\eta ^{\prime }$, but the singlet meson in the planar approximation) is
included in the computation, one finds a surprising result of{\it \ a
flavor-independent sea}\cite{ehq}. Namely, the original $\overline{u}$-$%
\overline{d}$ asymmetry due to the asymmetric $\pi ^{\pm }$ fluctuations is
just cancelled by the corresponding asymmetry due to the coherent neutral
meson emissions. As a result, there is now an equal number of $u\overline{u}$
and $d\,\overline{d},$ as well as $s\overline{s}$ pairs.

Mathematically, this flavor independence comes about as follows. Equating
the coupling constants $g_8=g_0$ in the extended vertex,
\begin{equation}
{\cal L}_I=g_8\sum_{i=1}^8\overline{\,q}\lambda _i\phi _iq+\sqrt{\frac 23}g_0%
\overline{q}\eta ^{\prime }q  \label{couple2}
\end{equation}
and squaring the amplitude, one obtains the probability distribution of
\[
\left| \Psi \right| ^2\propto \sum_{i=1}^8\left( \overline{\,q}\lambda
_iq\right) \left( \overline{\,q}\lambda _iq\right) +\frac 23\left( \overline{%
\,q}q\right) \left( \overline{\,q}q\right)
\]
which has the index structure as
\[
\sum_{i=1}^8\left( \,\lambda _i\right) _{ab}\left( \,\lambda _i\right) _{cd}+%
\frac 23\delta _{ab}\delta _{cd}=2\delta _{ad}\delta _{bc}.
\]
The r.h.s., deduced from a well-known identity of the Gell-Mann matrices\cite
{clbook}, clearly shows the flavor independent nature of the result.

While the mathematics of this flavor asymmetry cancellation is fairly
straightforward, the physics is more intriguing. It shows that the deviation
from an SU(3) symmetric sea should, for the most part, spring from the
nonplanar contributions. That nonplanar contributions are important for an
adequate description of the nonperturbative QCD is to be expected. In fact
the axial anomaly vanishes at the planar-diagram level. The resolution of
the $\eta ^{\prime }$ mass problem depends on the nonplanar physics\cite
{u1-vene}. However, with the admission of the nonplanar contributions, the
equality between the octet and singlet couplings is broken: $\,g_0/g_8\equiv
\zeta \neq 1\,\,\,.$We can now repeat the above calculation and express our
result in terms of the two parameters--- this coupling ratio $\zeta $ and $%
a, $ the probability of $\pi ^{+}$ emission: $\,u=2+\overline{u},\,\,d=1+%
\overline{d},\,\,\,$and$\,\,s=\overline{s},\,\,\,$with the antiquark
contents being:
\begin{eqnarray}
\overline{u} &=&\frac a3\left( \zeta ^2+2\zeta +6\right) ,\,\,\,\,\overline{d%
}=\frac a3\left( \zeta ^2+8\right) ,\,\,  \label{flav} \\
\,\,\overline{s} &=&\frac a3\left( \zeta ^2-2\zeta +10\right) .
\end{eqnarray}

Let us review our phenomenological knowledge of the flavor content of the
nucleon: The deviation from the Gottfried sum rule for the deep inelastic
lepton-nucleon scatterings is interpreted as showing an asymmetry between
the $\overline{u}$ and $\overline{d}$ quarks in the nucleon sea:

\begin{equation}
\left[ \int_0^1dx\frac{F_2^p(x)-F_2^n(x)}x-\frac 13\right] =\frac 23\left(
\overline{u}-\overline{d}\right) .  \label{grs}
\end{equation}
The NMC measurements did indeed show the Gottfried integral being
significantly different from one third:
\begin{equation}
I_G=0.235\pm 0.026\,\,.  \label{u-d}
\end{equation}
Here we have quoted the new New Muon Collaboration result published this year%
\cite{nmc}.

This $\overline{u}$-$\overline{d}$ asymmetry has now been confirmed by the
NA51 Collaboration\cite{na51} at CERN in a Drell-Yan experiment of
scattering protons on proton and on deuteron targets\cite{es-dy}. The
measured ratio of muon pair production cross sections $\sigma _{pp}$ and $%
\sigma _{pn\text{ }}$can be expressed as the antiquark content ratio:
\begin{equation}
\overline{u}/\overline{d}=0.51\pm 0.04\,(stat)\pm 0.05\,(syst).  \label{DY}
\end{equation}
This quantity is of particular interest in our model calculation, since it
depends only on one parameter:
\begin{equation}
\overline{u}/\overline{d}=\frac{\zeta ^2+2\zeta +6}{\zeta ^2+8}\,\,.
\label{rth}
\end{equation}
Interestingly, with only the octet Goldstone contribution (thus $\zeta =0$)
this ratio is fixed to be $0.75,$ comparing rather poorly with the
experimental result in Eq. (\ref{DY}). With the inclusion of the singlet
contribution ($\zeta \neq 0$), this expression still implies a lower bound
for the $\overline{u}/\overline{d}$ ratio of $1/2\,\,$at $\zeta =-2$, and an
upper bound of $5/3$ at $\zeta =4.$ More relevantly, the experimental value
in Eq. (\ref{DY}) implies that the coupling ratio $\zeta $ must be {\it %
negative: }$-4.3\leq \zeta \leq -0.7\,.$ Given the crudity of our
calculation and the sensitivity of the quadratic relations, we shall merely
illustrate our model calculation results in the subsequent discussion with
the following simple parameter choice:
\begin{equation}
a=0.1\,\,,\,\,\,\,\,\,\,\,\,\,\,\,\,\,\,\,\,\,\,\text{and \thinspace
\thinspace \thinspace \thinspace \thinspace \thinspace \thinspace \thinspace
\thinspace \thinspace \thinspace \thinspace \thinspace \thinspace \thinspace
\thinspace \thinspace \thinspace }\zeta =-1.2\,\,\,  \label{para}
\end{equation}
which corresponds to $\overline{u}/\overline{d}\,=0.53$, and to fixing $a$
in Eq. (\ref{flav}) so that Eq. (\ref{grs}) will yield the central
experimental value of the Gottfried sum:
\[
I_G=\frac 13+\frac 23\left( \overline{u}-\overline{d}\right) =\frac 13+\frac
49\left( \zeta -1\right) a=0.235\,\,\,.
\]

The fractions of quark flavors $\,f_a\equiv \frac{q_a+\overline{q_a}}{\sum
\left( q+\overline{q}\right) }\,\,\,$in the proton$\,\,$can also be
calculated from Eq. (\ref{flav}) with the parameters of Eq. (\ref{para}):
\begin{equation}
f_u\simeq 0.48,\,\,\,\,\,\,\,\,\,\,f_d\simeq
0.33,\,\,\,\,\,\,\,\,\,\,f_s\simeq 0.19.  \label{frth}
\end{equation}
Observationally, the strange quark fraction $f_{s\text{ }}$can be deduced
from the phenomenological value\cite{sigma} of the $\pi N$ sigma term
\begin{equation}
\sigma _{\pi N}=\widehat{m}\left\langle N\left| \overline{u}u+\overline{d}%
d\right| N\right\rangle =45\pm 10\,MeV  \label{sigma}
\end{equation}
where $\widehat{m}\equiv \frac 12\left( m_u+m_d\right) ,$ and the SU(3)
symmetry relation
\begin{equation}
\begin{array}{c}
M_8\equiv \frac 13\left( \widehat{m}-m_s\right) \left\langle N\left|
\overline{u}u+\overline{d}d-2\overline{s}s\right| N\right\rangle  \\
=M_\Lambda -M_\Xi \simeq -200\,MeV
\end{array}
.  \label{m8}
\end{equation}
Using the current algebra result of $m_s/\widehat{m}=25$ we obtain the ratio
\[
1-2y\equiv \frac{\left\langle \overline{u}u+\overline{d}d-2\overline{s}%
s\right\rangle }{\left\langle \overline{u}u+\overline{d}d\right\rangle }=%
\frac{3\widehat{m}M_8}{\left( \widehat{m}-m_s\right) \sigma _{\pi N}}\simeq
\frac 59.
\]
Keeping in mind that the scalar operator $\overline{q}q$ measures the sum of
the quark and antiquark numbers (as opposed to $q^{\dagger }q,$ which
measures the difference), we find
\begin{equation}
\left( f_s\right) _{\sigma _{\pi N}}=\frac y{1+y}\simeq 0.18\,\,\,.\,\,
\label{frexp}
\end{equation}
The good agreement between this and the value in Eqs. (\ref{frth}) should be
regarded as fortuitous in view of the fact the value in Eq. (\ref{frexp})
involves the flavor SU(3) symmetry relation (\ref{m8}). Thus an uncertainty
of at least 20\%, above and beyond that shown in Eq. (\ref{sigma}), has to
be included in this estimate.

We now turn to the nucleon spin content. In the no interaction limit, the
spin-up proton ($p_{+}$) wave function $\,p_{+}=\frac 1{\sqrt{6}}\left(
2u_{+}u_{+}d_{-}-u_{+}u_{-}d_{+}-u_{-}u_{+}d_{+}\right) \,$ implies that the
probabilities of finding $u_{+}$ ($u$ quark in the spin-up state), $u_{-},\,$
$d_{+}$ and $d_{-}$ in $p_{+}$ are $\frac 53,\,\frac 13$, $\frac 13$ and $%
\frac 23$, respectively. These values will be altered by the same meson
emission processes of the chiral quark model as discussed above in
connection with proton's flavor content. The total probability of Goldstone
emission being $\frac a3\left( \zeta ^2+8\right) $, the contributions of the
various spin states, after one interaction, can then be read off from
\[
\begin{array}{c}
\left[ 1-\frac a3\left( \zeta ^2+8\right) \right] \left( \frac 53u_{+}+\frac
13u_{-}+\frac 13d_{+}+\frac 23d_{-}\right)  \\
+\frac 53\left| \Psi \left( u_{+}\right) \right| ^2+\frac 13\left| \Psi
\left( u_{-}\right) \right| ^2+\frac 13\left| \Psi \left( d_{+}\right)
\right| ^2+\frac 23\left| \Psi \left( d_{-}\right) \right| ^2,
\end{array}
\]
where amplitudes $\Psi \left( q_{\pm }\right) \,$can be calculated in an
entirely similar manner as that done for Eq. (\ref{psi2}). The quark
contributions to the proton spin $\,\Delta q=q_{+}-q_{-}\,$ are:
\begin{eqnarray}
\Delta u &=&\frac 43-\frac 19\left( 8\zeta ^2+37\right) a,  \label{spin} \\
\,\,\,\Delta d &=&-\frac 13+\frac 29\left( \zeta ^2-1\right) a, \\
\,\,\Delta s &=&-a
\end{eqnarray}
which for the parameters of Eq. (\ref{para}) yields the values:
\begin{equation}
\begin{array}{c}
\Delta u=0.79,\,\,\,\,\,\,\,\,\,\,\,\,\,\Delta
d=-0.32,\,\,\,\,\,\,\,\,\,\,\,\Delta s=-0.10.
\end{array}
\label{spinth}
\end{equation}

Since the 1988 announcement by EMC of their proton spin content result,
there has been significant new experimental developments in the measurement
of the spin-dependent structure functions of the neutron, as well as that of
the proton\cite{nuspin}. In the meantime further higher order perturbative
QCD calculations have been carried out for such spin-dependent structure
function sum rules\cite{qcdc}. Taking into account of these higher order
results [hence the perturbative QCD predicted $Q^2$ dependence through $%
\alpha _s(Q^2)$], Ellis and Karliner have recently shown\cite{ellisk} that
all the diverse experimental measurements are consistent with each other,
and that the fundamental Bjorken sum rule is verified to about 12\%. When
the final result, after using a flavor SU(3) symmetry relation [similar to
that of Eq. (\ref{m8})], is expressed in terms of the individual quark
contributions to the proton spin,
\begin{equation}
\Delta u_{\exp }=0.83,\,\,\Delta d_{\exp }=-0.42,\,\,\Delta s_{\exp }=-0.10,
\label{spin}
\end{equation}
we have a comparison with our result in Eq. (\ref{spinth}). One of the
principal outcome of the this recent round of experimentation and
phenomenological analysis is the conclusion that the total quark spin
contribution actually does not vanish: \thinspace $\Delta \Sigma \equiv
\Delta u+\Delta d+\Delta s\simeq 0.31$, which is to be compared to our model
calculation result of $0.37$ in Eq. (\ref{spinth}).

We should also note that ours is an SU(3) {\it symmetric} computation. The
basic feature that the strange quark is heavier than the up and down quarks
have not been taken into account. The meson emission corrections for each
component of an SU(3) multiplet must, in such a calculation, be the same.
Consequently, the naive quark model ratio of $\Delta _3$/$\Delta _8=$ $5/3$
\thinspace is unchanged (where $\Delta _3=$ $\Delta u-\Delta d$ and $\Delta
_8=\Delta u+\Delta d-2\Delta s).$ It is about $25\%$ lower than the
phenomenological value\cite{f/d} of $\left( \Delta _3/\Delta _8\right)
_{\exp }\simeq $ $2.1.$ And, the similarly defined flavor-fraction ratio of $%
f_3/f_8=1/3$ is greater than the experimental value of $\left(
f_3/f_8\right) _{\exp }\simeq 0.23.$ Inclusion of the SU(3) breakings will
necessarily increase the number of parameters and complicate the model
calculation. We postpone such an endeavor to a later stage, and choose to
present our principal results without having them masked by this
complication.

Overall we find it rather encouraging that this simple calculation in a
theoretically well-motivated framework can account for many of the puzzling
features discovered in recent years of the spin and flavor contents of the
nucleon. To us what is significant is this broad pattern of agreement in one
unified calculation. Our effort overlaps with many of the previous works\cite
{prev}, where these effects have been discussed as unrelated phenomena. What
we wish to emphasize in this work is that the nonrelativistic quark model,
when supplemented with the Goldstone structure, does yield an adequate
approximation of the observed low energy physics\cite{close}. A key
ingredient in this implementation is the inclusion of the ninth Goldstone
boson with a differently renormalized coupling $g_0\simeq -1.2g_8$.
Presumably the success with such an inclusion shows that, above the
confinement scale, the nonplanar mechanism which endows the $\eta ^{\prime }$
with a mass can still be treated as a perturbation as suggested by the 1/N$%
_c $ expansion of QCD.

With this first encouraging result, it might be worthwhile to embark on a
more elaborate field theoretical calculation. This will involve more mass,
cutoff parameters, and further phenomenological inputs, but it will also
allow a more detailed comparison of the $Q^2$ and $x$ dependences of the
structure functions between the chiral quark theory expectations and the
experimental measurements.\smallskip\

One of us (L.F.L.) would like thank Ernest Henley for useful conversation
and for bringing Ref.\cite{na51} to our attention. This work is supported in
at CMU by the Department of Energy (DE-FG02-91ER-40682), and at UM-St.Louis
by an RIA award and by the National Science Foundation (PHY-9207026).

\begin{itemize}
\item  E-mail addresses: {\sc stcheng@slvaxa.umsl.edu}, and {\sc %
lfli@bethe.phys.cmu.edu}${\cal \ }$

\item  {\bf Figure Caption: }Fig. 1. Spin and flavor corrections due to
Goldstone boson fluctuations. (a) A spin zero meson couples to quarks of
opposite helicites. (b) Production of a $(q^{\prime }\overline{q^{\prime }})$
pair via Goldstone emission.
\end{itemize}

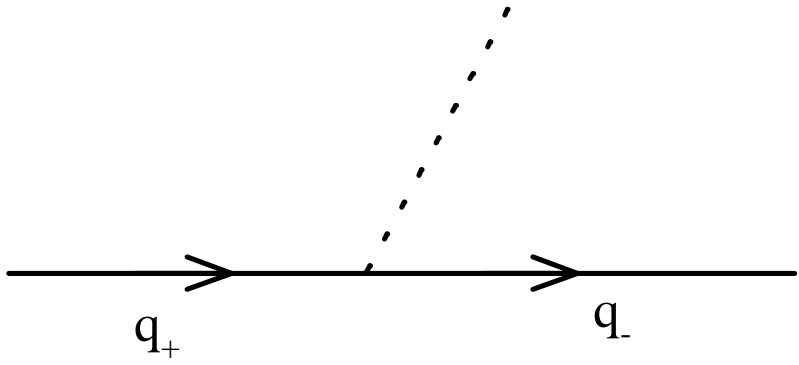
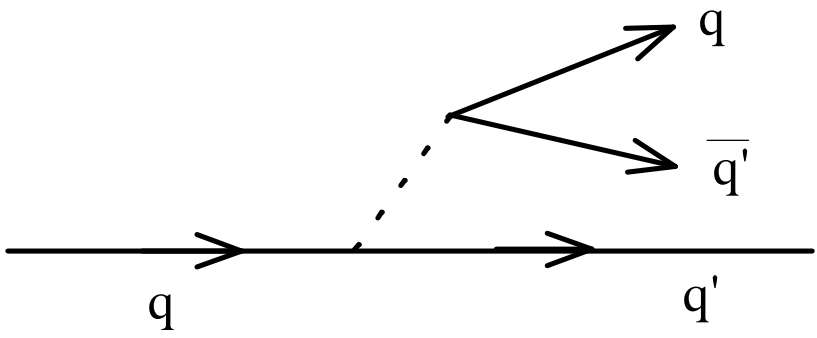
\end{document}